\documentclass[12pt,,letterpaper]{JHEP3}
\usepackage{graphicx}
\usepackage{epsfig}

\title{Time drift of subtended angles as a new cosmological probe}

\author{Hongbao Zhang\\
Crete Center for Theoretical Physics, Department of Physics, \\
University of Crete, 71003 Heraklion, Greece\\
\email{hzhang@physics.uoc.gr} }
\author{Zong-Hong Zhu\\
Department of Astronomy, Beijing Normal University, Beijing, 100875,
China \email{zhuzh@bnu.edu.cn}} \preprint{CCTP-2009-17}
 \abstract{We here propose the time drift of subtended angles as a
 new possible
cosmological probe. In particular, with the coming era of
microarcsecond astrometry, our proposal can be used to measure the
Hubble expansion rate of our universe in a direct way.}
\begin{document}
\section{Introduction}
Research in cosmology has become extraordinarily lively in the past
quarter century. In particular, the use of Type Ia supernovae as
standard candles lead to the discovery that the expansion of our
universe is accelerating, implying that the energy of the universe
may be dominated by some sort of exotic matter, with a ratio of
pressure to density less than one third, dubbed dark
energy\cite{Weinberg}. However, besides the cosmological constant,
there are various other dark energy models devised to explain the
mysterious accelerating expansion. On the other hand, the current
accelerating expansion of our universe can also be accounted for by
either the modification of Einstein gravity or the violation of
Copernican principle. With the other cosmological probes, some of
them has been ruled out, but there still remain a number of
theoretical models surviving such observational tests\cite{FTH}.

\begin{figure}
\includegraphics[width=12cm,height=10cm]
{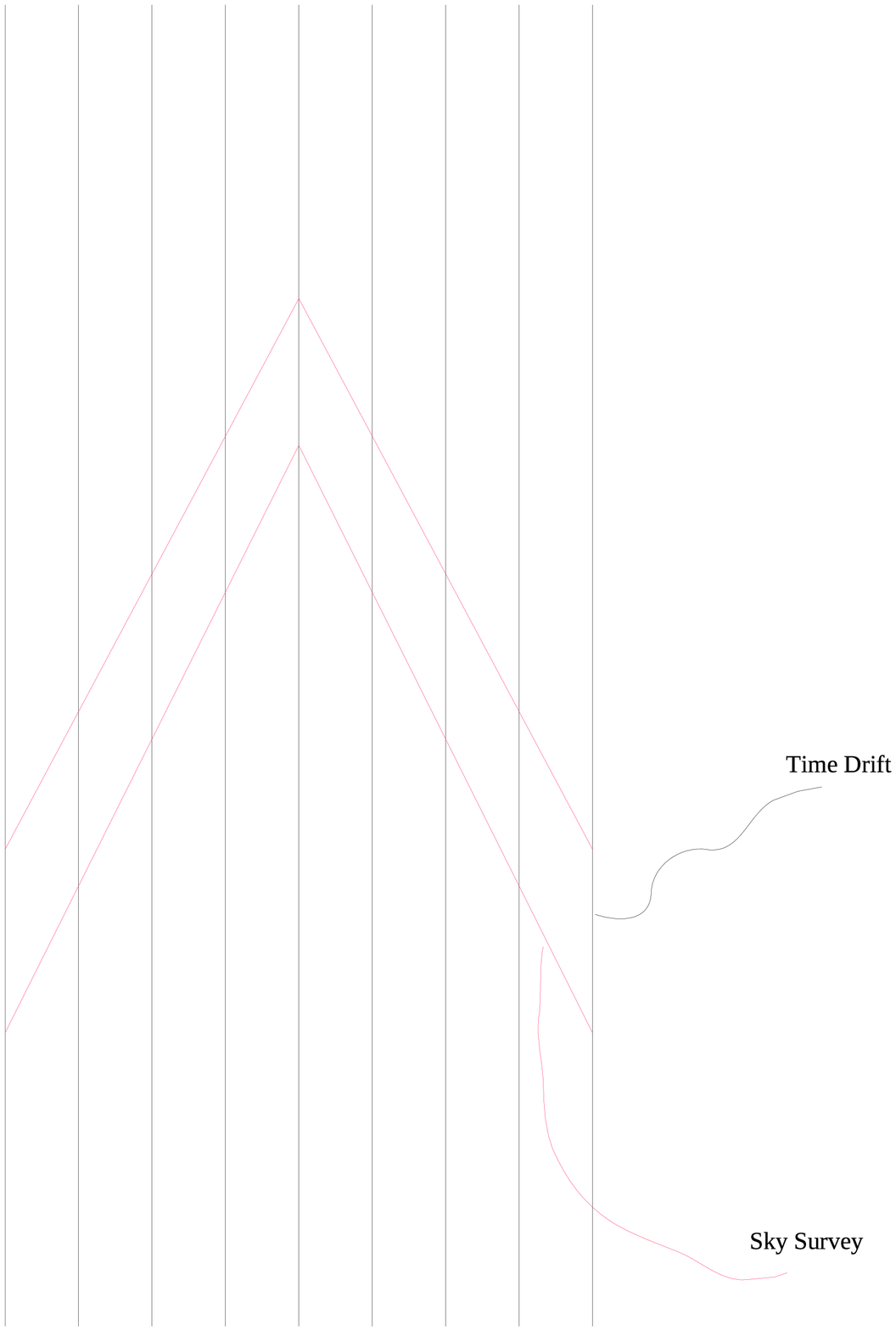}\\
  \caption{Time drift versus sky survey in the redshift space.}\label{td}
\end{figure}

In this era of empirical cosmology, it is thus significant to
propose some new observational programs that may shed light on our
understanding of the universe in a way independent from other known
cosmological probes. By comparing and combining results from very
different methods of determining cosmological parameters, we may
both obtain stronger constraints than any method alone would impose,
and test techniques against one another to identify signatures of
systematic effects. In particular, in a field so afflicted by
systematic errors as cosmology, having many independent but
complementary techniques is the best way to ensure that we are on
the right track to explore the genuine mechanism underlying the
evolution of our universe. We here contribute to such a theme by
proposing the time drift of subtended angles as a new cosmological
probe. As depicted in Fig.\ref{td}, the ordinary cosmological probes
are usually related to the sky survey along the past light cone of
today. However, we have another way to acquire the evolution
information of our universe by measuring the time drift of
cosmological objects. The classical exemplification is Sandage-Loeb
test\cite{Sandage,Loeb}.

In next section we shall popularize the spirit of time drift by
confining ourselves onto the case of subtended angles and argue that
the time drift of subtended angle may provide us with a new
cosmological probe, just like Sandage-Loeb test. Some discussions
will be presented in the last section.
\section{Time drift of subtended angles}
In what follows we assume that on large scales our universe is
described by the FLRW metric
\begin{equation}
ds^2=-dt^2+a^2(t)(\frac{dr^2}{1-kr^2}+r^2d\Omega^2),
\end{equation}
where $k=1,0,-1$ correspond to closed, flat, and open universes,
respectively.

Imagine a luminous object that extends a proper distance $D_\perp$
perpendicular to the line of sight. Suppose the light was emitted
from the object at the time $t_1$ and is observed by us at $t_0$. By
the definition of the angular diameter distance, the object will
subtend an angle as follows
\begin{equation}
\theta=\frac{D_\perp}{d_A(z)},\label{angle}
\end{equation}
where
\begin{equation}
d_A(z)=a(t_1)S[\int_{t_1}^{t_0}\frac{dt}{a(t)}]=\frac{a(t_0)}{1+z}S[\frac{1}{a(t_0)}\int_0^z\frac{dz'}{H(z')}]
\end{equation}
with $S[x]=\sin x, x,\sinh x$ for closed, flat, and open universes,
respectively.

Then after a time integral $\delta t_0$, the variation of the
subtended angle gives
\begin{eqnarray}
\delta\theta&=&-\frac{D_\perp}{d^2_A(z)}\delta d_A(z)+\frac{\delta D_\perp}{d_A(z)}\nonumber\\
&=&-\frac{D_\perp}{d^2_A(z)}\{\dot{a}(t_1)\delta
t_1S[\int_{t_1}^{t_0}\frac{dt}{a(t)}]+a(t_1)S'[\int_{t_1}^{t_0}\frac{dt}{a(t)}](\frac{\delta
t_0}{a(t_0)}-\frac{\delta t_1}{a(t_1)})\}+\frac{\dot{D}_\perp\delta t_1}{d_A(z)}\nonumber\\
&=&-\frac{D_\perp}{d_A(z)}[H(z)-\frac{\dot{D}_\perp}{D_\perp}]\delta
t_1=-\frac{H(z)-\frac{\dot{D}_\perp}{D_\perp}}{1+z}\theta\delta
t_0,\label{coin}
\end{eqnarray}
where $S'$ denotes the differentiation with respect to its argument,
and $\frac{\delta t_0}{\delta t_1}=\frac{a(t_0)}{a(t_1)}=1+z$ has
been used. If the size evolution rate of the luminous object can be
ignored compared with the Hubble flow, then the time drift of the
subtended angle can be expressed as
\begin{equation}
\delta v\equiv|\frac{\delta\theta}{\theta}|=\frac{H(z)}{1+z}\delta
t_0,\label{resolution}
\end{equation}
which implies that in principle we can determine $H(z)$ by
measurement of the subtended angle at the redshift $z$ and its
variation over the time interval $\delta t_0$.

Now let us make a rough estimate of the feasibility of such a
proposal by specializing to the fiducial concordance $\Lambda$CDM
model, where the Hubble expansion rate is given by
\begin{equation}
H(z)=H_0\sqrt{\Omega_m^0(1+z)^3+1-\Omega_m^0}\label{cdm}
\end{equation}
with the Hubble constant $H_0=70kms^{-1}Mpc^{-1}$ and
$\Omega_m^0=0.3$.
\begin{figure}
\includegraphics
{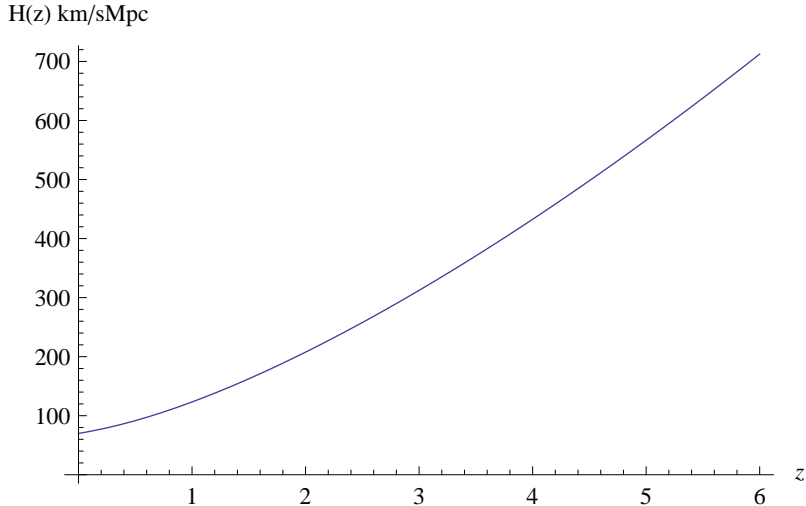}\\
  \caption{Hubble expansion rate as a function of redshift in the concordance $\Lambda$CDM model.}\label{hp}
\end{figure}

By Eq.(\ref{cdm}), the Hubble expansion rate increases with
redshift, in particular, as shown in Fig.\ref{hp}, goes up from
$70kms^{-1}Mpc^{-1}$ at the present to $700kms^{-1}Mpc^{-1}$ or so
at the redshift $z=6$. It is difficult to imagine that the magnitude
of  $\frac{\dot{D}_\perp}{\dot{D}}$ can be achieved to the same
order as the Hubble expansion rate for those virialized galaxies or
clusters of galaxies. Thus at least such virialized systems can
serve as those cosmological objects for us to apply our proposal to.

\begin{figure}
\includegraphics
{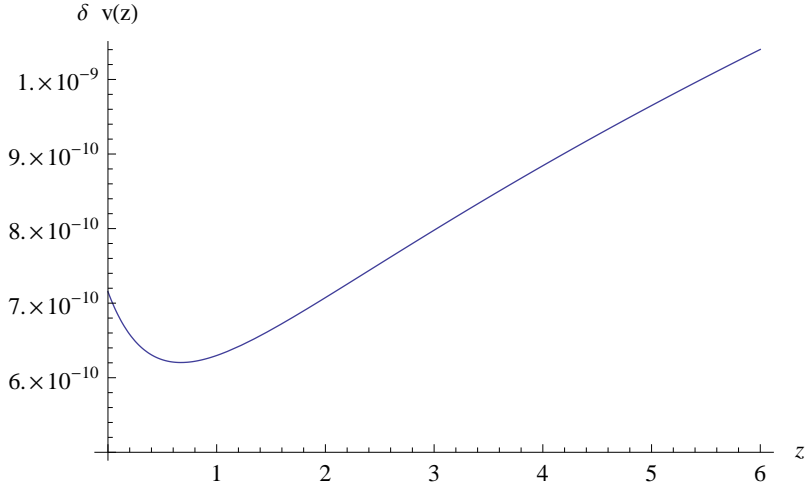}\\
  \caption{Time drift of subtended angles as a function of redshift in the concordance $\Lambda$CDM
  model, where the time baseline takes ten years.}\label{tdsa}
\end{figure}

\begin{figure}
\includegraphics
{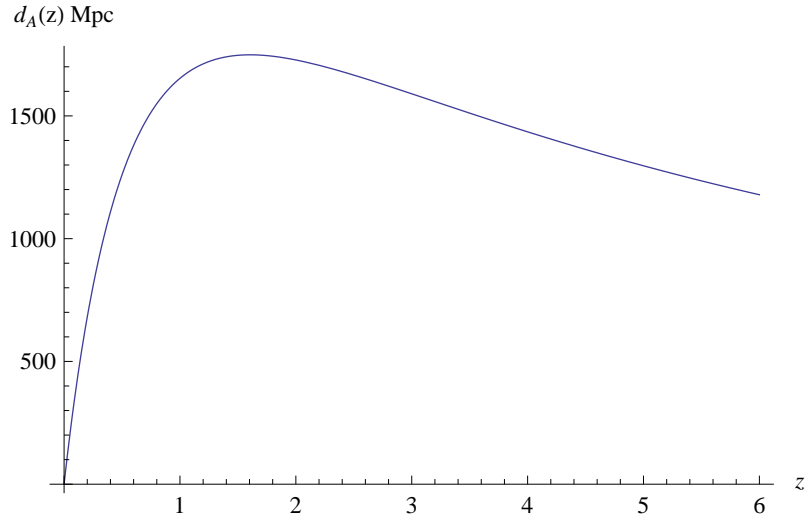}\\
  \caption{Angular diameter distance as a function of redshift in the concordance $\Lambda$CDM
  model.}\label{add}
\end{figure}

Before proceeding, let us recall the fact that we have been getting
ready for an era in which the microarcsecond astrometry will become
norm\cite{Brown}. Thus if we assume a ten year baseline for $\delta
t_0$, then as shown in Fig.\ref{tdsa}, it follows from
Eq.(\ref{resolution}) that the subtended angle of the observed
object should be at least one tenth degree such that its variation
can be detected in the microarcsecond telescopes. If we start from
the nearby object with the angle diameter distance $100Mpc$, its
size requires to be of order $0.1Mpc$, which can actually be
satisfied by some galaxies, not to mention the typical clusters of
galaxies. Furthermore, as illustrated in Fig.\ref{add}, contrary to
our naive intuition, the angular diameter distance decreases at high
redshifts. In particular, it arrive at its maximum of around
$2000Mpc$ at the redshift $z=1.5$ or so. This implies that the
maximal size of the required luminous object is of order $1Mpc$,
which is the typical size of clusters of galaxies. Therefore, the
time drift of subtended angle seems to be a very promising
cosmological probe.

\section{Discussions}
We have argued that the time drift of subtended angles may be used
as a new promising cosmological probe to measure the Hubble
expansion rate of our universe in a direct way.

However, so far the analysis of its feasibility rests on an order of
magnitude estimate. It is thus important to see whether the
realistic data will distort such an estimate dramatically. On the
other hand, although it is hard to imagine that the virialized
systems will undertake a size evolution to the same order as the
Hubble expansion, it may be expected that at least a few of them
will violate this general belief. Nevertheless we may employ the
other side of the same coin by instead using Eq.(\ref{coin}) to
determine the size evolution rate for those exceptional objects and
non-virialized ones if we know the evolution law of our universe
from other cosmological probes, and vice versa, since the subtended
angle by definition entangles the whole universe with those luminous
objects in it.

It is interesting to ask whether we can disentangle them
intrinsically. One way is to combine it with the time drift of the
redshift difference from the far and near points of the object to
obtain an evolution free cosmological probe, reminiscent of
Alcock-Paczynski test\cite{AP}. Another way out is to go directly
for the time drift of angle diameter distance, which by expression
is immune to the size evolution effect totally.

All of these issues are worthy of further investigation but beyond
the scope of this paper, we expect to report them elsewhere in the
near future.
\section*{Acknowledgements}
We are grateful to Vassilis Charmandaris, Lulu Fan, Andrea Lapi,
Carlo Nipoti, Tomonori Totani, Steven Weinberg, and Pengjie Zhang
for very helpful correspondence on the size evolution problem. We
are also much indebted to Scott Dodelson for his stimulating
discussions, which sharpen the feasibility issue of our proposal
greatly. HZ was partially supported by a European Union grant
FP7-REGPOT-2008-1-CreteHEP Cosmo-228644 and a CNRS PICS grant \#
4172. ZHZ was supported by the National Science Foundation of China
under the Distinguished Young Scholar Grant 10825313, the Key
Project Grant 10533010, and by the Ministry of Science and
Technology National Basic Science Program (Project 973) under grant
\# 2007CB815401.

\end{document}